\documentclass[aps,prd,twocolumn,superscriptaddress,amsfont,amssymb,amsmath,nofootinbib,showpacs,balancelastpage]{revtex4-1}

\usepackage{graphicx,longtable,natbib,mathrsfs,xcolor,ulem,array,float}
\usepackage{txfonts,subfigure,aas_macros}

\newcommand{\bs}{\boldsymbol}

\newcommand{\Msun}{M_\odot}

\begin{document}

\addtolength{\hoffset}{-0.525cm}
\addtolength{\textwidth}{1.05cm}
\title{Spin speed correlations and the evolution of galaxy-halo systems}

\author{Ming-Jie Sheng}
\affiliation{Department of Astronomy, Xiamen University, Xiamen, Fujian 361005, China}

\author{Lin Zhu}
\affiliation{Department of Astronomy, Xiamen University, Xiamen, Fujian 361005, China}

\author{Hao-Ran Yu}\email{haoran@xmu.edu.cn}
\affiliation{Department of Astronomy, Xiamen University, Xiamen, Fujian 361005, China}

\author{Hong-Chuan Ma}
\affiliation{Department of Astronomy, Xiamen University, Xiamen, Fujian 361005, China}

\author{Hai-Kun Li}
\affiliation{Department of Astronomy, Xiamen University, Xiamen, Fujian 361005, China}

\author{Peng Wang}
\affiliation{Shanghai Astronomical Observatory, Chinese Academy of Sciences, Shanghai 200030, China}
\affiliation{Astronomical Research Center, Shanghai Science and Technology Museum, Shanghai, 201306, China}

\author{Xi Kang}
\affiliation{Institute for Astronomy, the School of Physics, Zhejiang University, Hangzhou 310027, China}
\affiliation{Center for Cosmology and Computational Astrophysics, Zhejiang University, Hangzhou 310027, China}
\affiliation{Purple Mountain Observatory, 10 Yuan Hua Road, Nanjing 210034, China}

\date{\today}

\begin{abstract}
Galaxy angular momenta (spins) contain valuable cosmological information, complementing their positions and velocities. The baryonic spin direction of galaxies has been probed as a reliable tracer of their host halos and the primordial spin modes. Here we use the TNG100 simulation of the IllustrisTNG project to study the spin magnitude correlations between dark matter, gas, and stellar components of galaxy-halo systems and their evolutions across cosmic history. We find that these components generate similar initial spin magnitudes from the same tidal torque in Lagrangian space. At low redshifts, the gas component still traces the spin magnitude of the dark matter halo and the primordial spin magnitude. However, the traceability of the stellar component depends on the $ex$ $situ$ stellar mass fraction, $f_{\rm acc}$. Our results suggest that the galaxy baryonic spin magnitude can also serve as a tracer of their host halo and the initial perturbations, and the galaxy-halo correlations are affected by the similarity of their evolution histories.
\end{abstract}

\maketitle

\section{Introduction} \label{sec.intro}
In cosmology, the low-redshift large-scale structure (LSS) of the universe evolved from the 
primordial density fluctuations of the early universe. One of the key tasks of the LSS study is 
looking for a link between the cosmic initial conditions and low-redshift observables
\citep{2005MNRAS.360L..82R,2021JCAP...06..024M}.
In general, the LSS is primarily driven by the dynamics of dark matter (DM). After
recombination, baryonic matter decouples from radiation and follows the clustering of DM 
under gravity. Hence, the matter distribution on a large scale can be probed by various 
tracers, such as galaxies, resulting in rich cosmological information 
\citep{1969ApJ...155..393P}.

The locations and peculiar velocities of galaxies are traditionally used as the tracers 
of the LSS to probe the primordial perturbations, while the rotations of galaxies provide another 
degree of freedom to extract additional cosmological information. At low redshifts, the 
three-dimensional (3D) angular momenta (spins) of the galaxy are observable via their ellipticities, 
projection angles, spiral parities, Doppler effects, and dust absorptions \citep{2019ApJ...886..133I}.
The tidal torque theory explains the generation of the angular momentum of a clustering system 
in {\it Lagrangian space}\footnote{The Lagrangian space is defined as the {\it initial}, {\it comoving} 
coordinates of mass elements in the picture of structure formation. In $N$-body simulations, $N$-body 
particles represent phase-space ``sheets'' \cite[][Chapter 12]{2020moco.book.....D}, and when the 
initial conditions of the simulation are set at sufficiently high redshift, their comoving coordinates 
represent Lagrangian space.} \citep{1969ApJ...155..393P,1970Afz.....6..581D,1984ApJ...286...38W}. 
The tidal torque, generated by the misalignment between the moment of inertia of 
protohalos/protogalaxies (DM halos/galaxies in Lagrangian space) and the tidal fields they feel, 
provides a direction-invariant and persistent generation of angular momentum until the virialization 
of halos. These virialized DM halos at low redshifts tend to keep the predicted angular momentum 
directions \citep{2000ApJ...532L...5L,2002MNRAS.332..339P} and magnitudes 
\citep{2021PhRvD.103f3522W}. Thus, their angular momenta provide independent cosmological information,
including, e.g., the reconstruction of primordial density and tidal fields
\citep{2000ApJ...532L...5L,2001ApJ...555..106L},
the effects of cosmic neutrino mass \citep{2019PhRvD..99l3532Y,2020ApJ...898L..27L} and
dark energy \citep{2020ApJ...902...22L}, the possible detection of chiral
violation \citep{2020PhRvL.124j1302Y,2022PhRvD.105h3512M},
and the understanding of galaxy intrinsic alignments \citep{2001MNRAS.320L...7CF, 
2011JCAP...05..010B,2015JCAP...10..032S,2018MNRAS.473.1562W}. 

Practically, the rotations of DM halos are difficult to observe, so we can only expect the angular 
momenta of galaxies or other baryonic tracers to be the proxies of those of the DM halos. 
\citep{2021NatAs...5..283M}, for the first time, discovered a weak but significant correlation 
between the observational galaxy spins and the cosmic initial conditions. Most recently, 
\citep{2023ApJ...943..128S} found that the baryonic components of galaxies trace 
the spin directions of their host DM halos and the primordial spin modes in the IllustrisTNG-100 simulations.
The highly nonlinear baryonic effects, including gas cooling, galaxy and star formation, and supernova and 
black hole feedbacks, have not fully erased the memory of the initial spin directions. However, the 
studies on the correlations of spin magnitude between galaxies and their host halos have not reached 
full agreement. \citep{2016MNRAS.460.4466Z} found a strong correlation between the evolution of 
the specific angular momenta (sAM) of galaxy baryonic components and DM halo in the EAGLE simulations 
\citep{2015MNRAS.450.1937C,2015MNRAS.446..521S}.
\citep{2022ApJ...937L..18D} similarly showed that the overall angular momentum is retained in a nearly 
constant ratio during star formation and gas circulation in the IllustrisTNG-50 simulations. 
Observationally, \cite{2023MNRAS.518.1002R} suggested that galaxies with larger baryon fractions have 
also retained larger fractions of their sAM in the process of galaxy formation and evolution.
However, \citep{2019MNRAS.488.4801J} found almost no correlation between the spin parameters of galaxies 
and their host halos using the VELA \citep{2014MNRAS.442.1545C,2015MNRAS.450.2327Z} and NIHAO 
\citep{2015MNRAS.454...83W} zoom-in simulations. 
In addition, it is also unclear whether and how baryonic components trace the primordial spin magnitude 
across cosmic evolution. 
In this work, we use the state-of-the-art magneto-hydrodynamical (MHD) simulations IllustrisTNG 
\citep{2018MNRAS.480.5113M,2018MNRAS.475..624N, 2019MNRAS.490.3234N,2018MNRAS.475..648P,
2019MNRAS.490.3196P,2018MNRAS.475..676S,2019ComAC...6....2N} to study the spin magnitude 
correlations, characterized by the kinematic spin speed and supportedness, between the baryonic 
components of galaxies and their host DM halos. We will further investigate how the primordial 
spin magnitude can be traced by baryonic matter at low redshifts.

This paper is organized as follows. In Sec. \ref{sec.meth}, we briefly describe the 
simulation and analytical methods. Sec. \ref{sec.resu} shows the spin magnitude 
correlation and evolution results for galaxy-halo systems.
The conclusion and discussion are presented in Sec. \ref{sec.conclu}.

\section{Methodology}\label{sec.meth}
\subsection{TNG100 simulation and galaxy samples}
The IllustrisTNG simulations are a suite of MHD galaxy formation simulations using the 
{\small AREPO} code \citep{2010MNRAS.401..791S,2020ApJS..248...32W}. 
In this study, the main results are given by the TNG100-1 simulation, which starts with 
$1820^3$ DM particles and $1820^3$ gas cells in a periodic cubic box with a comoving 
length of $75 \,h^{-1} {\rm Mpc}$ per side. 
The initial condition is generated with the Zel'dovich approximation and the N-G{\small EN}IC 
code \citep{2015ascl.soft02003S}. 
The adopted cosmological parameters are from the Plank 2015 results \citep{2016A&A...594A..13P}, 
i.e., $\Omega_{\rm m}=0.3089$, $\Omega_{\rm b}=0.0486$, 
$\Omega_\Lambda=0.6911$, and $h=0.6774$. The mass resolutions for DM particles and gas 
cells are $m_{\rm DM}=7.5\times10^6\,\Msun$ and $m_{\rm gas}=
1.4\times10^6\,\Msun$ (on average), respectively.
DM halos and subhalos are identified with the friends-of-friends 
(FOF; \citep{1985ApJ...292..371D}) and {\small SUBFIND} algorithms 
\citep{2001MNRAS.328..726S}.
The TNG100-1 simulation has sufficient massive 
galaxy clusters, which enable us to study spin correlations on large, linear scales, while 
having a higher baryonic resolution compared with the TNG300-1 to study the spins of gas and stars.

In this paper, all the quantities of a galaxy and its host subhalo are calculated for the entire 
{\small SUBFIND} objects, i.e., using all particles belonging to these objects.
We consider only the central galaxies that belong to the most massive subhalos of 
their host halos with stellar masses $M_\ast \geq 10^{10}\,\Msun$, yielding a 
galaxy catalog that contains 3971 samples in the Eulerian space (redshift $z=0$) with particle IDs, 
positions, velocities, and other astrophysical properties for DM, gas, and stellar components, respectively.
To study the primordial spin mode, we need to trace the subhalo/galaxy mass elements back to the 
Lagrangian space (initial condition, redshift $z=127$). 
For DM, we can simply trace them by following the particle IDs. 
For gas cells and star particles, we trace their tracer particles \citep{2013MNRAS.435.1426G} back to 
the initial condition. The TNG100-1 simulation contains $2\times1820^3$ tracer particles. 

To quantify the galaxy morphology, we employ the kappa parameter, $\kappa_{\rm rot}$, which measures 
the fraction of the stellar kinetic energy invested into ordered rotation
\citep{2010MNRAS.409.1541S,2017MNRAS.467.3083R}.
It is defined as
\begin{equation}
  \kappa_{\rm rot} = \frac{K_{\rm rot}}{K}= \frac{1}{K}\sum_i\frac{1}{2}m_i\left(\frac{j_{z,i}}{R_i}\right)^2,
\end{equation}
where $K$ is the total kinetic energy of the stellar component, 
$m_i$ is the mass of a stellar particle, $j_{z,i}$ is the $z$-component of the particle's 
sAM, assuming that the $z$-axis coincides with the stellar angular 
momentum of the galaxy, $R_i$ is the particle's distance to the $z$-axis, and the sum
is carried out over all stellar particles in the galaxy. Following the 
classification in \citep{2012MNRAS.423.1544S}, galaxies with $\kappa_{\rm rot}<0.5$ and 
$\kappa_{\rm rot}\geq0.7$ are referred to as spheroid- or disc-dominated, respectively.
The former morphology contains 1308 galaxies in our samples; the second contains 507 
galaxies. The remainder consist of intermediate types where both rotation and velocity 
dispersion play comparable structural roles. 

\subsection{Spin parameters}
In Eulerian (Lagrangian) space, the angular momentum vector $\bs J_{\rm E}$ ($\bs J_{\rm L}$) of a 
certain subhalo/galaxy component (e.g., DM, gas, or stars) is computed as
\begin{eqnarray} 
  \bs J_{\rm E}&=&\sum_i m_i \bs x_i' \times \bs v_i',\\
  \bs J_{\rm L}&=&\sum_i m_i \bs q_i' \times \bs u_i',
\end{eqnarray}
where $\bs x_i'=\bs x_i - \bar{\bs x}$, $\bs v_i'=\bs v_i - \bar{\bs v}$, 
$\bs q_i'=\bs q_i - \bar{\bs q}$, $\bs u_i'=\bs u_i - \bar{\bs u}$, with
$m_i$, $\bs x_i$ ($\bs q_i$) and $\bs v_i$ ($\bs u_i$) are the particle mass,  
Eulerian (Lagrangian\footnote{We refer the readers to \citep[][Chapter 12]{2020moco.book.....D} for more details 
on the definitions of Lagrangian properties.}) 
position, and velocity of the $i{\rm th}$ particle, while $\bar{\bs x}$ 
($\bar{\bs q}$) and $\bar{\bs v}$ ($\bar{\bs u}$) are the Eulerian (Lagrangian) center-of-mass position 
and mean velocity of this component.
Then the sAM vector is defined as $\bs j=\bs J/\sum_i m_i$. \citep{2022ApJ...937L..18D} showed that the 
sAM of gas ($j_{\rm g}$) and stellar ($j_{\rm s}$) components almost conserved that of the dark matter
halo ($j_{\rm h}$) during star formation and gas circulation in the central disc-dominated galaxies 
($\kappa_{\rm rot}\geq0.7$).

The Eulerian angular momentum of a virialized object can be characterized by the dimensionless 
spin parameter $\lambda_{\rm P} \equiv J\left\lvert E \right\rvert^{1/2}G^{-1}M^{-5/2}$ 
\citep{1969ApJ...155..393P} or the closely related definition $\lambda_{\rm B} 
\equiv J/(\sqrt{2}MVR)$ \citep{2001ApJ...555..240B},
where $J$, $E$, $M$, $V$, $R$ are the total angular momentum, total energy, mass, circular 
velocity, radius of the system, and $G$ is the Newton's constant. The parameters $\lambda_{\rm P}$ 
and $\lambda_{\rm B}$ are very similar for typical NFW halos \citep{2001ApJ...555..240B}.
Previous work using the VELA and NIHAO zoom-in simulations found a null correlation between the 
spin parameter $\lambda_{\rm B}$ of galaxies ($\lambda_{\rm gal}$) and their host halos 
($\lambda_{\rm halo}$), especially at redshift $z \geq 1$ \citep{2019MNRAS.488.4801J}. 

The dimension of sAM gives
\begin{equation}
  {\rm dim}(\bs j)=\frac{\rm{M \cdot L^2T^{-1}}}{\rm M}=\rm{L^2T^{-1}},
\end{equation}
which shows that sAM is sensitive to the system size, especially when comparing different 
components and galaxies in different mass ranges. 
Besides, not all the above spin parameters are straightforwardly defined for the baryonic components 
and for protohalos/protogalaxies in the Lagrangian space. Here, we employ two parameters according 
to the kinematics of mass elements to characterize the rotation of different components for a 
subhalo/galaxy in both Eulerian and Lagrangian spaces to avoid these problems. 

\begin{figure*}[htbp]
  \centering    
  \subfigure{\includegraphics[width=0.62\columnwidth]{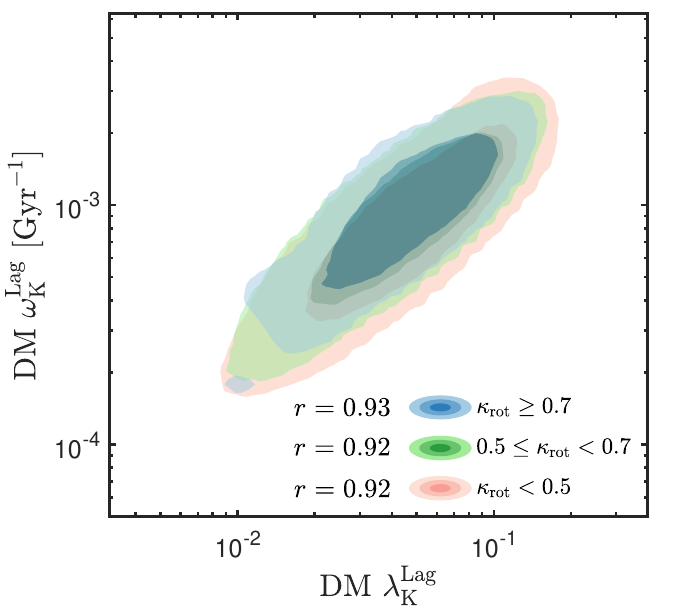}}
  \subfigure{\includegraphics[width=0.62\columnwidth]{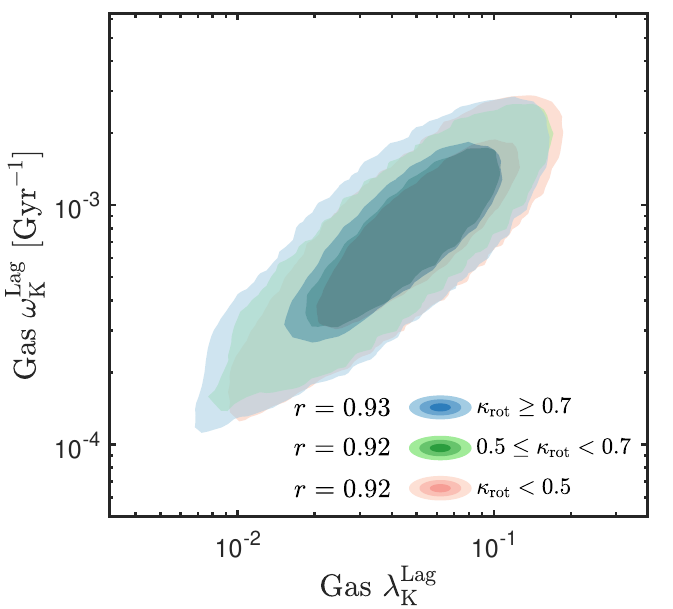}}
  \subfigure{\includegraphics[width=0.62\columnwidth]{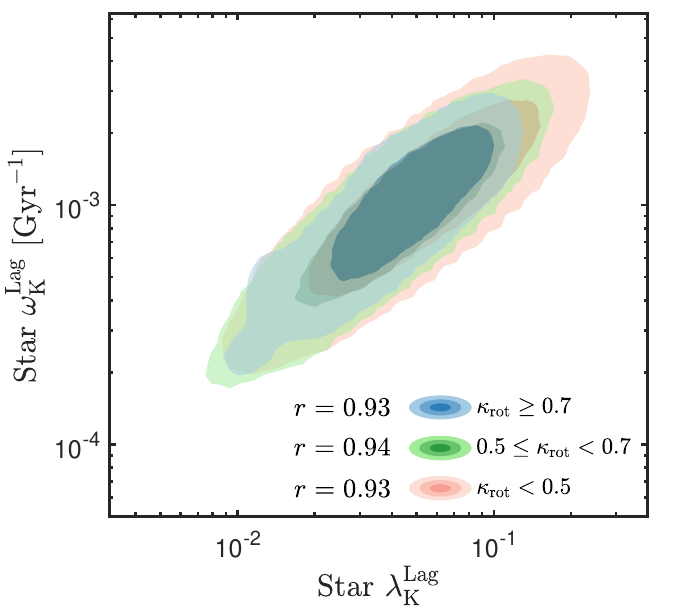}}
  \subfigure{\includegraphics[width=0.62\columnwidth]{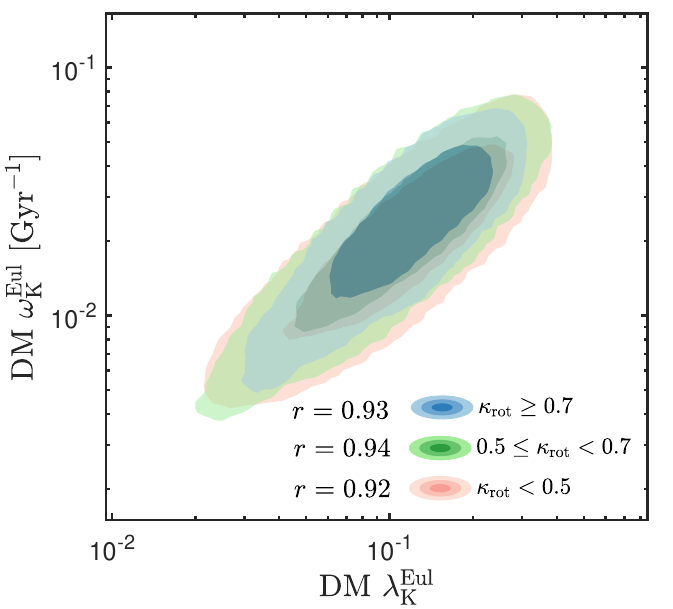}}
  \subfigure{\includegraphics[width=0.62\columnwidth]{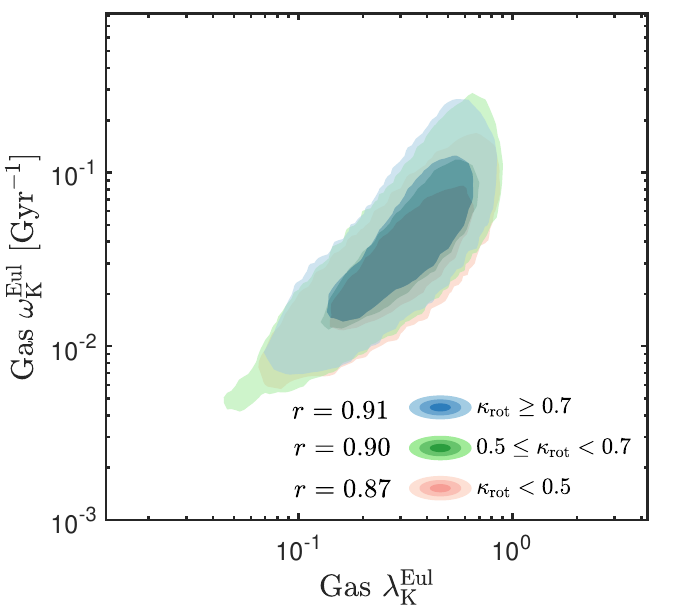}}
  \subfigure{\includegraphics[width=0.62\columnwidth]{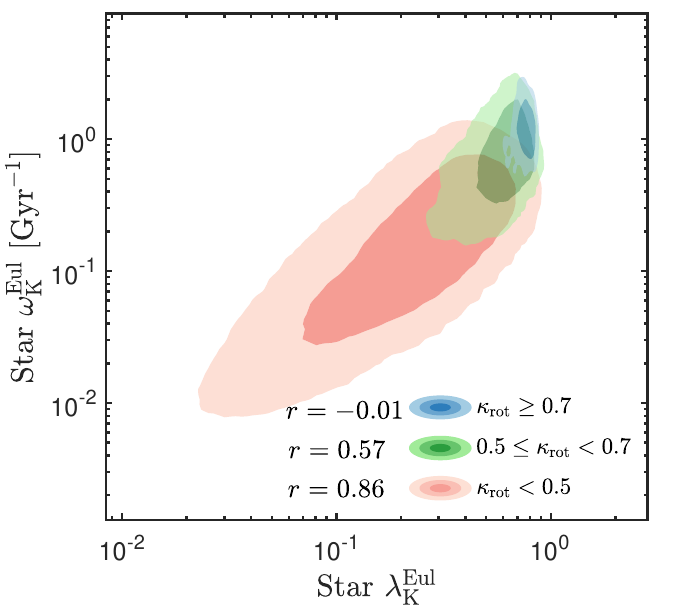}}
  \caption{Comparison of average spin speed $\omega_{\rm K}$ and spin supportedness 
  $\lambda_{\rm K}$ for DM (left column), gas (middle column), and stars (right column) 
  in Lagrangian (upper row) and Eulerian (lower row) spaces, respectively. 
  The contours are colored according to $\kappa_{\rm rot}$, with the inner and outer regions 
  containing 68 and 95 percent of the galaxy population, respectively.
  The Pearson correlation coefficients $r$ (in logarithmic units) for different types of 
  galaxies are indicated in each panel.}
  \label{fig.1}
\end{figure*}

For a subhalo/galaxy that occupies region $V_x$ in Eulerian space and the corresponding 
region $V_q$ in Lagrangian space, the spin speed parameter is defined as 
\begin{eqnarray} 
  \omega_{\rm K}^{\rm Eul} &\equiv&
  \frac{\int_{V_x} \hat{j}_i \epsilon_{ijk} x'_j v'_k \,{\rm d}M}{2\pi\int_{V_x} r_i^2 \,{\rm d}M} \nonumber\\
  &=& 
  \frac{\int_{V_x} \sin\theta_1 \cos\theta_2 x'v' \,{\rm d}M}
  {2\pi\int_{V_x} \sin^2\theta_3x'^2  \,{\rm d}M}
  = \frac{J_{\rm E}}{I_{\hat{j}}},
\end{eqnarray}
where $\hat{j_i}=(\bs J_{\rm E}/J_{\rm E})_i$ is the unit $\bs J_{\rm E}$ vector ($J_E=|\bs J_{\rm E}|$), 
$x'=|\bs x'|$, $v'=|\bs v'|$, $\sin\theta_1=\sin{(\bs x',\bs v')}$, $\cos\theta_2=
\cos\left(\bs x' \times \bs v',\bs J_{\rm E}\right)$, $\sin\theta_3=\sin{(\bs x',
\bs J_{\rm E})}$, and $I_{\hat{j}}$ denotes the average moment of inertia along the spin direction.
This parameter can be similarly defined in Lagrangian space and denoted with 
$\omega_{\rm K}^{\rm Lag}$\footnote{Hereafter, the superscript $^{\rm Eul}$ denotes the spin 
parameters that are measured in Eulerian space, while the superscript $^{\rm Lag}$ denotes Lagrangian 
space.}. 

The spin supportedness parameter is defined as \citep{2021PhRvD.103f3522W,2022PhRvD.105f3540S}
\begin{equation}\label{eq.lambdaK}
  \lambda_{\rm K}^{\rm Eul} \equiv \frac{\int_{V_x} \hat{j}_i \epsilon_{ijk} x'_j v'_k \,{\rm d}M}
  {\int_{V_x} x' v' \,{\rm d}M}=
  \frac{\int_{V_x} \sin\theta_1 \cos\theta_2 x' v' \,{\rm d}M}{\int_{V_x} x' v' \,{\rm d}M},
\end{equation}
which can also be similarly defined in Lagrangian space and denoted with $\lambda_{\rm K}^{\rm Lag}$. 

These two parameters characterize the spin magnitude of a subhalo/galaxy from similar but subtly different 
perspectives.
$\omega_{\rm K}$ characterizes the average spin speed of the system along the spin
direction with dimension $\dim\,(\omega_{\rm K})={\rm time}^{-1}$, in the unit of ${\rm Gyr}^{-1}$, which 
is independent of the mass and size of the system.
On the other hand, $\lambda_{\rm K} \in [0, 1]$ is dimensionless with $\dim\,(\lambda_{\rm K})=1$ and 
characterizes whether the system is rotation-supported ($\lambda_{\rm K} \rightarrow 1$) or 
dispersion-supported ($\lambda_{\rm K} \rightarrow 0$). 
For instance, a coplanar system with all mass elements having homodromous circular orbits has 
$\lambda_{\rm K}=1$, and a rotating rigid isodensity globe has $\lambda_{\rm K} = 8/3\pi \simeq 0.85$,
while their spin speed $\omega_{\rm K}$ may be varied. We can directly study the spin magnitude 
of different components both in Lagrangian and Eulerian space by utilizing these parameters, regardless of 
the mass and size of the galaxies. In addition, combining these two parameters gives us an idea of how 
fast or slow and to what extent the galaxy is rotating as a whole.

\begin{figure}[htbp]
  \centering    
  \subfigure{\includegraphics[width=0.49\columnwidth]{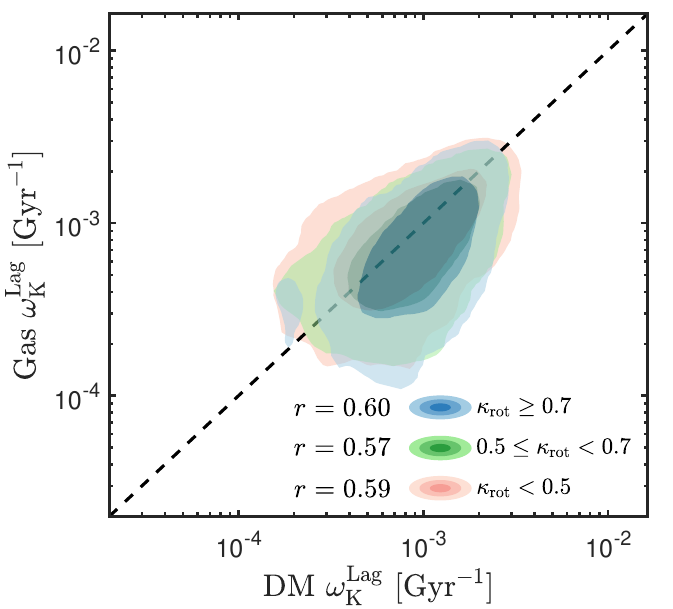}}
  \subfigure{\includegraphics[width=0.49\columnwidth]{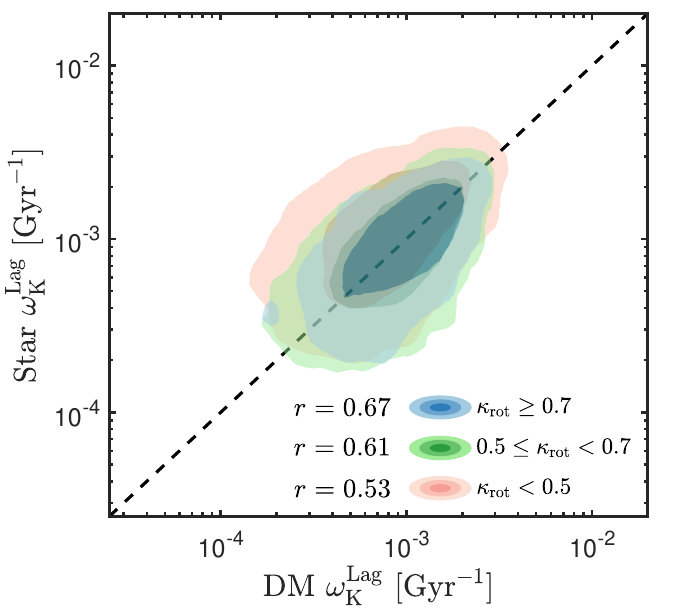}}
  \subfigure{\includegraphics[width=0.49\columnwidth]{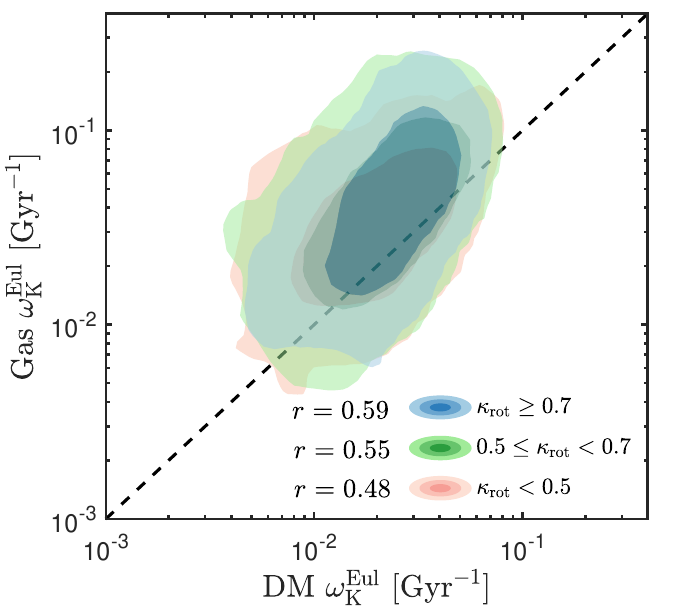}}
  \subfigure{\includegraphics[width=0.49\columnwidth]{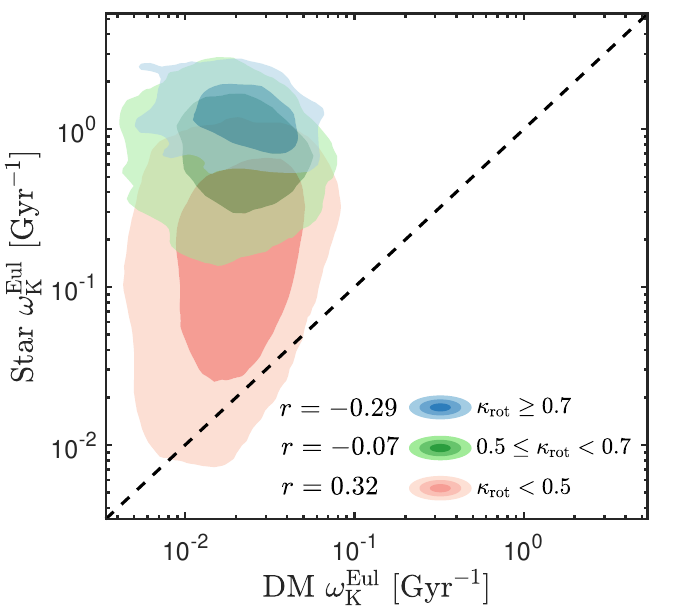}}
  \caption{Comparison of average spin speed $\omega_{\rm K}$ between DM and gas (left column) / star (right column) 
  in the Lagrangian (upper row) and Eulerian (lower row) spaces, respectively. The contours are colored according to 
  $\kappa_{\rm rot}$, with the inner and outer regions containing 68 and 95 percent of the galaxy population, 
  respectively. The Pearson correlation coefficients $r$ for different types of galaxies are indicated in 
  each panel. The dotted lines indicate $y=x$.}
  \label{fig.2}
\end{figure}

\begin{figure}[htbp]
  \centering    
  \subfigure{\includegraphics[width=0.49\columnwidth]{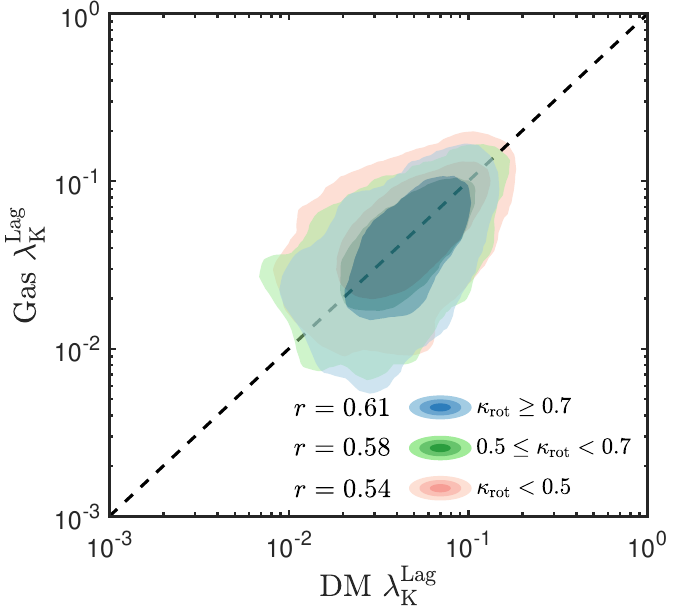}}
  \subfigure{\includegraphics[width=0.49\columnwidth]{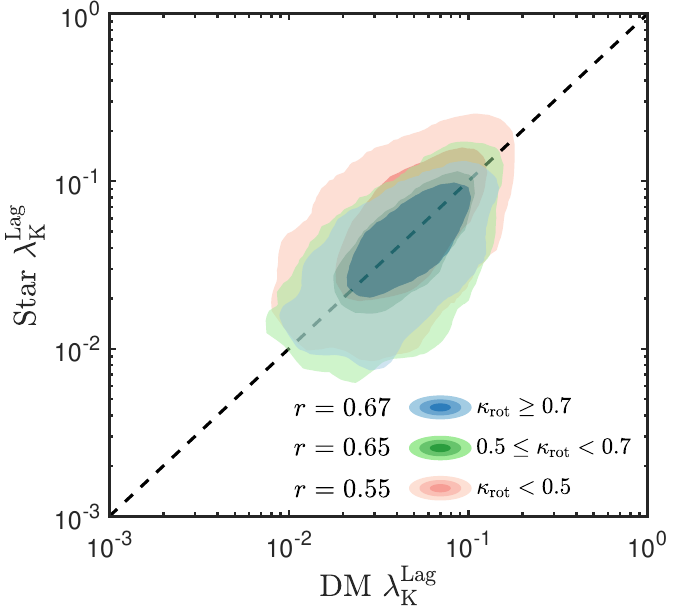}}
  \subfigure{\includegraphics[width=0.49\columnwidth]{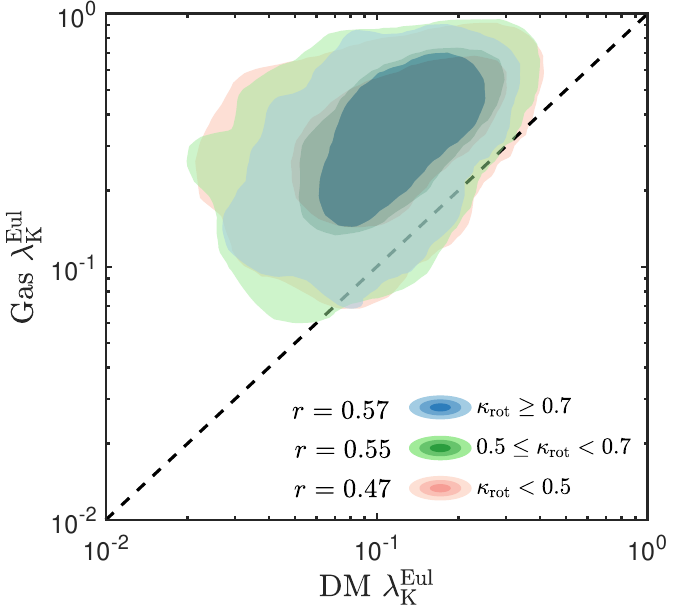}}
  \subfigure{\includegraphics[width=0.49\columnwidth]{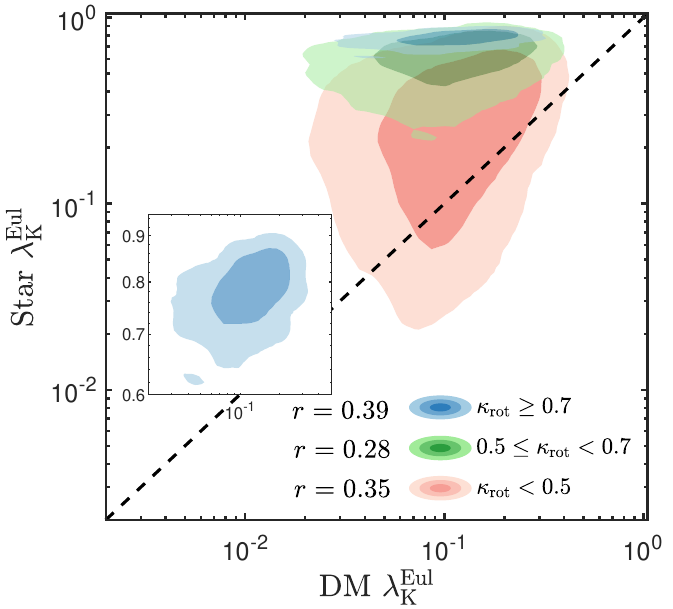}}
  \caption{Similar plotted as Fig. \ref{fig.2}, except for comparison of spin supportedness 
  $\lambda_{\rm K}$. The inset in the bottom-right panel shows the zoom-in contour of the 
  disc-dominated galaxies with $\kappa_{\rm rot}\geq 0.7$.}
  \label{fig.3}
\end{figure}

\begin{figure*}[htbp]
  \centering    
  \subfigure{\includegraphics[width=0.63\columnwidth]{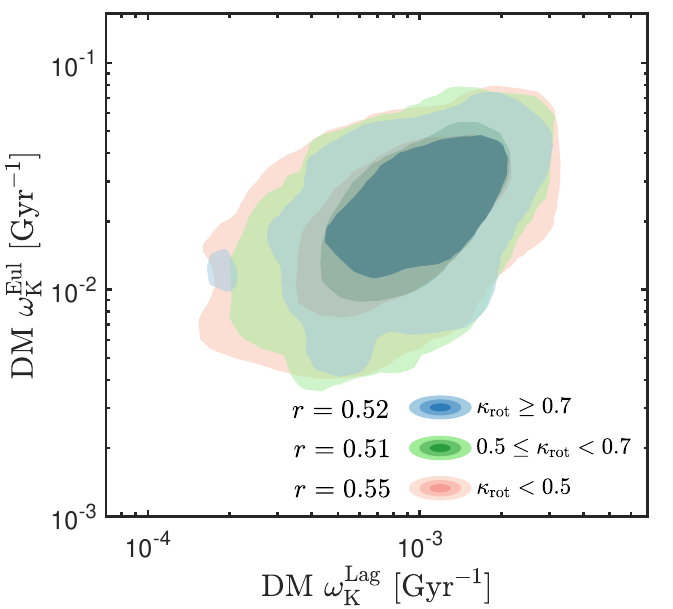}}
  \subfigure{\includegraphics[width=0.63\columnwidth]{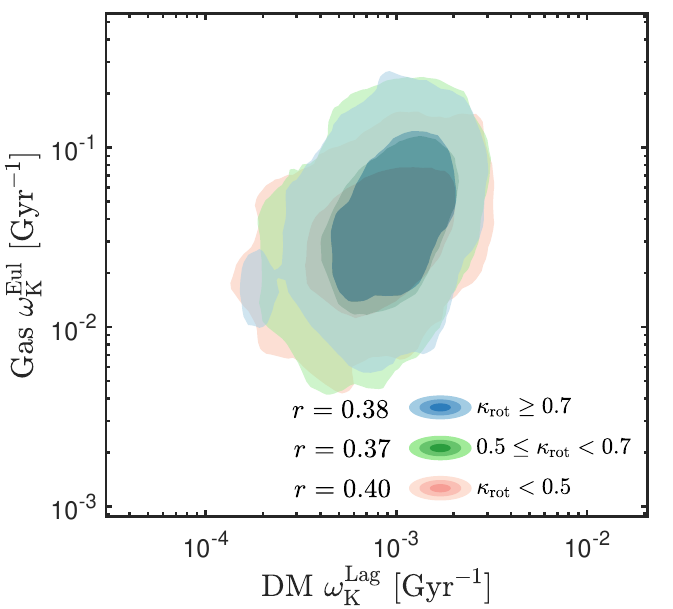}}
  \subfigure{\includegraphics[width=0.63\columnwidth]{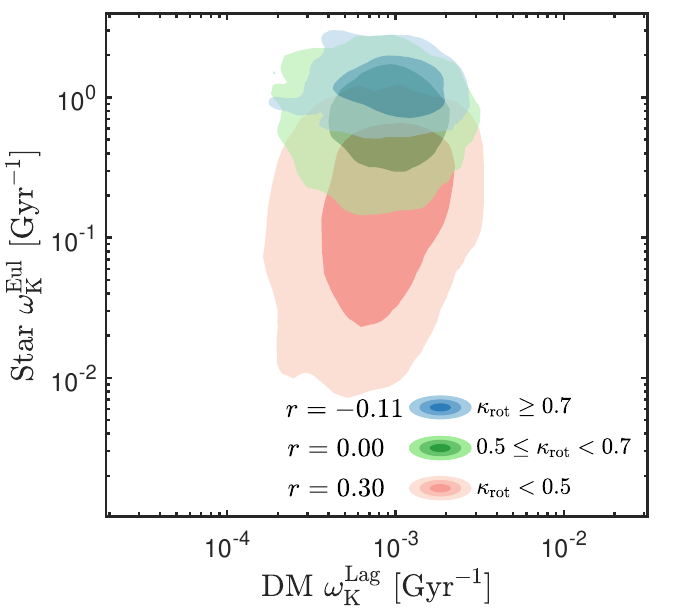}}
  \caption{Comparison of average spin speed $\omega_{\rm K}$ for DM (left column), gas (middle column), and stellar 
  (right column) components in the Eulerian space with the DM component in the Lagrangian space, respectively. The contours are 
  colored according to $\kappa_{\rm rot}$, with the inner and outer regions containing 68 and 95 percent of 
  the galaxy population, respectively. The Pearson correlation coefficients $r$ for different 
  types of galaxies are indicated in each panel.}
  \label{fig.4}
\end{figure*}

\begin{figure*}[htbp]
  \centering    
  \subfigure{\includegraphics[width=0.63\columnwidth]{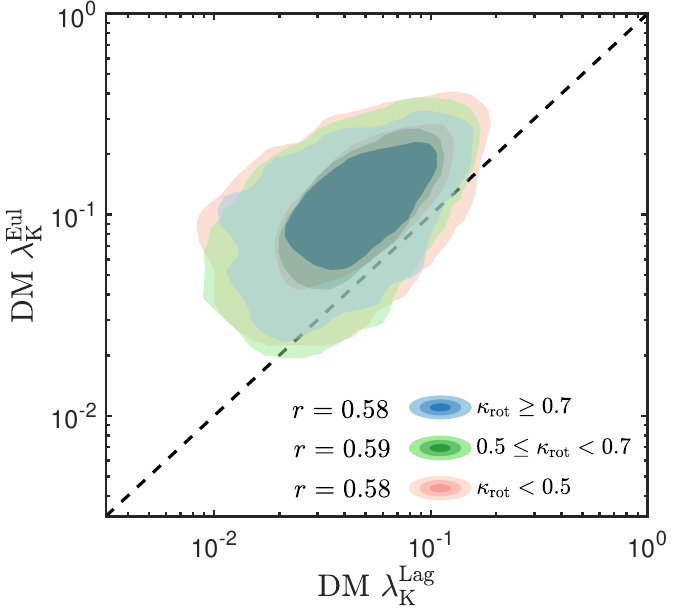}}
  \subfigure{\includegraphics[width=0.63\columnwidth]{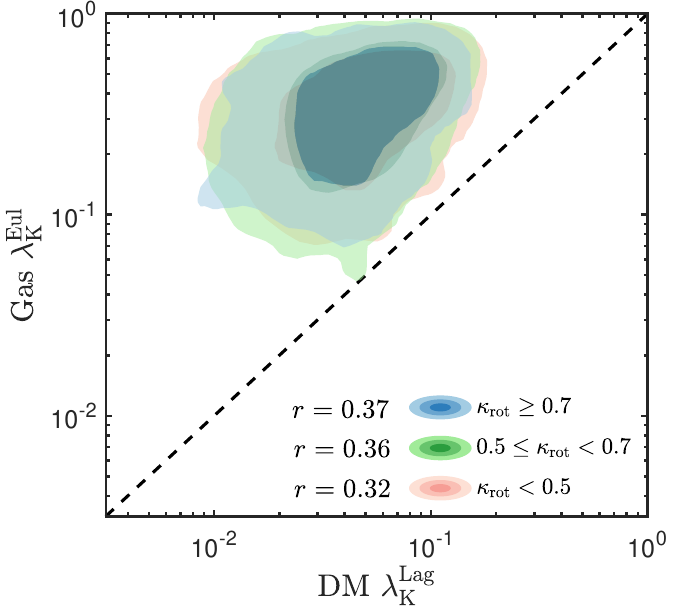}}
  \subfigure{\includegraphics[width=0.63\columnwidth]{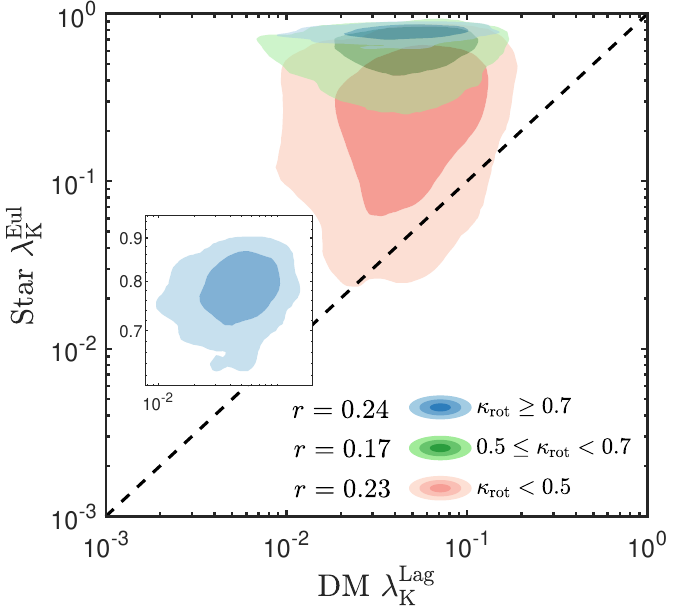}}
  \caption{Similar plotted as Fig. \ref{fig.4}, except for comparison of spin supportedness 
  $\lambda_{\rm K}$. The dotted lines indicate $y=x$. The inset in the rightmost panel shows 
  the zoom-in contour of the disc-dominated galaxies with $\kappa_{\rm rot}\geq 0.7$.}
  \label{fig.5}
\end{figure*}

\section{Results}\label{sec.resu}
\subsection{Spin speed-supportedness correlation}\label{sec.3.1}
We first start with a comparison between the galaxy-halo average spin speed and 
spin supportedness. Unless otherwise noted, we will color disc-dominated, intermediate-type, and 
spheroid-dominated galaxies with blue, green, and red in the rest of this paper.
The upper panels of Fig. \ref{fig.1} show that the average spin speed and 
spin supportedness of DM, gas, and stellar components have similar high correlations in 
Lagrangian space, with the Pearson correlation coefficients $r$ (hereafter calculated in 
logarithmic units) larger than $0.9$, which is independent of galaxy morphology. 
The lower panels also show strong correlations for each component in Eulerian space. 
But the bottom-right panel indicates that the spin speed and supportedness 
of the stellar component are clearly dependent on galaxy morphology. This can be attributed to the similar 
kinematic definitions of $\omega_{\rm K}$ and $\kappa_{\rm rot}$ when considering stellar component, 
both of which describe the spin supportedness of stars along the spin direction. 
In addition, the stellar angular momenta of disc-dominated galaxies are mainly distributed in the 
inner region of their host halos, while a large fraction of the stellar angular momenta of spheroid-dominated 
galaxies are found beyond two stellar half-mass radii \citep{2017MNRAS.467.3083R}. Due to the inward 
distribution of angular momenta and higher spin supportedness, the average spin speed of the stellar 
component of disc-dominated galaxies is much higher than that of spheroid-dominated galaxies. 

\subsection{Spin magnitude of different components}\label{sec.3.2}
In this subsection we compare the spin magnitude between DM, gas, and stellar components 
in Lagrangian and Eulerian spaces, respectively. 
In the upper panels of Fig. \ref{fig.2}, we show the average spin speeds characterized by 
$\omega_{\rm K}$ for different components in Lagrangian space. Clearly, the average spin speeds of 
gas and stellar components are strongly correlated with the DM component and match the $y=x$ line in 
Lagrangian space, which is independent of galaxy morphology. In the upper panels of Fig. \ref{fig.3}, 
we find similar results for spin supportedness $\lambda_{\rm K}$. These similar correlations could 
be explained by the fact that these components have similar mass distributions and feel the same tidal 
torque in Lagrangian space \citep{2023ApJ...943..128S}.

In Eulerian space, both the bottom-left panels of Fig. \ref{fig.2} and Fig. \ref{fig.3} show that 
the $\omega_{\rm K}$ and $\lambda_{\rm K}$ of gas are still correlated with DM, regardless of galaxy 
morphology. The spin speed of gas is slightly higher than that of DM, while the spin supportedness 
of gas far exceeds. 
However, the bottom-right panel of Fig. \ref{fig.2} 
suggests that the spin speed of the stellar component is poorly correlated with the DM component, with a 
Pearson correlation coefficient of $r=0.19$, especially for disc-dominated and intermediate-type galaxies. 
As mentioned above, disc-dominated galaxies have much higher stellar spin speeds than 
spheroid-dominated ones. In addition, the spin speed of the stellar component is higher than that of the DM 
component, especially for disc-dominated galaxies. The correlation of spin supportedness between DM 
and stellar components is slightly higher with the Pearson correlation coefficient $r=0.23$, shown in the 
bottom-right panel of Fig. \ref{fig.3}. Considering only the disc-dominated galaxies, the correlation 
coefficient increases to $r=0.39$. Similarly, disc-dominated galaxies also have much higher spin 
supportedness than spheroid-dominated ones. Although the spin direction of the stellar component 
is still correlated with the host halo in Eulerian space and the initial direction in Lagrangian space shown 
in \citep{2023ApJ...943..128S}, these correlations vanish when it comes to the spin magnitude.
Meanwhile, the poor correlations of spin speed and supportedness between the stellar and DM components in Eulerian space 
are consistent with \citep{2019MNRAS.488.4801J}, who found an almost null correlation of spin parameter $\lambda_{\rm B}$ 
between the total halos and the galaxies in the inner region. We will explore the possible origins of these poor 
correlations in Sec. \ref{sec.3.4}.

\subsection{Evolution of spin magnitude}\label{sec.3.3}
Then we focus on the evolution of galaxy-halo spin magnitude between the Lagrangian and 
Eulerian spaces. In Fig. \ref{fig.4} and Fig. \ref{fig.5}, we compare the average spin speed 
$\omega_{\rm K}$ and spin supportedness $\lambda_{\rm K}$ for different components of galaxy-halo 
systems with their original protohalos. Clearly, the spin speed and supportedness increase for 
each component through cosmic evolution, which is expected by the tidal torque theory. 
In addition, the leftmost and middle panels of Fig. \ref{fig.4} and Fig. \ref{fig.5} suggest that 
the spin speed and supportedness of DM and gas components are both correlated with the Lagrangian 
protohalos, which is independent of galaxy morphology. For protohalos that acquire higher spin magnitude 
from the initial tidal field, the DM and gas components of the final halos tend to have higher spin speed and 
supportedness.
The gas component shows weaker correlation with the Pearson correlation coefficients $r=0.39$ for 
$\omega_{\rm K}$ and $r=0.35$ for $\lambda_{\rm K}$ than the DM component, with $r=0.54$ for $\omega_{\rm K}$ 
and $r=0.59$ for $\lambda_{\rm K}$. These decreases can be partially explained by the effects of baryonic 
processes \citep{2017ApJ...841...16D}, especially the stellar and AGN feedback 
\citep{2017MNRAS.466.1625Z}. The DM and gas components of galaxy-halo systems still retain the 
memory of the initial spin magnitude of their host halos, which is similarly found for spin directions in 
\citep{2023ApJ...943..128S}.
However, for the stellar component, the rightmost panel of Fig. \ref{fig.4} shows that the spin speed 
is poorly correlated with the protohalos, especially for disc-dominated and intermediate-type galaxies. The 
rightmost panel of Fig. \ref{fig.5} shows similar a weak correlation for spin supportedness. 

\subsection{Origin of the weak star-DM correlation}\label{sec.3.4}
In the previous subsections, we find that the spin magnitude of the gas component well 
traces the DM component both in Lagrangian and Eulerian space, but does not apply to stars. Here we show 
that the spin magnitude correlation between the stellar and DM components depends on the $ex$ $situ$ 
stellar mass fraction, $f_{\rm acc}$, which measures the fraction of a galaxy's stellar mass contributed 
by stars that formed in other galaxies and which were subsequently accreted.
This quantity is provided in the stellar assembly catalogs of TNG100-1 \citep{2015MNRAS.449...49R,
2016MNRAS.458.2371R,2017MNRAS.467.3083R}.

\begin{figure*}[htbp]
  \centering    
  \subfigure{\includegraphics[width=0.63\columnwidth]{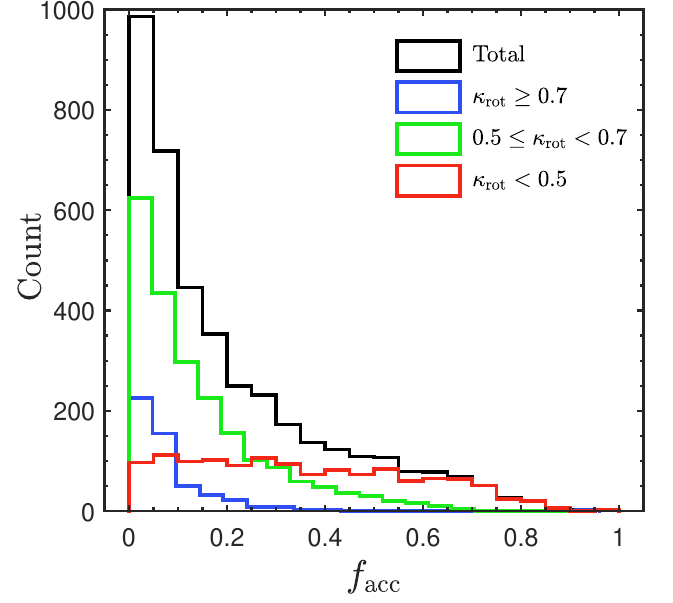}}
  \subfigure{\includegraphics[width=0.63\columnwidth]{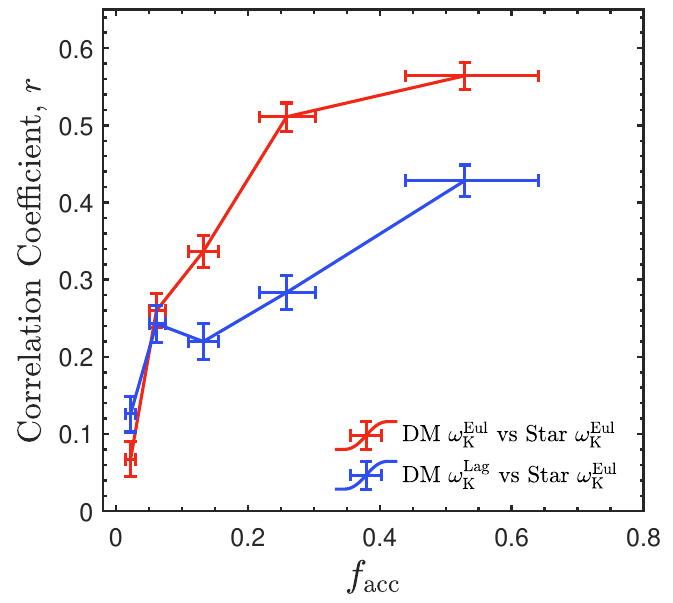}}
  \subfigure{\includegraphics[width=0.63\columnwidth]{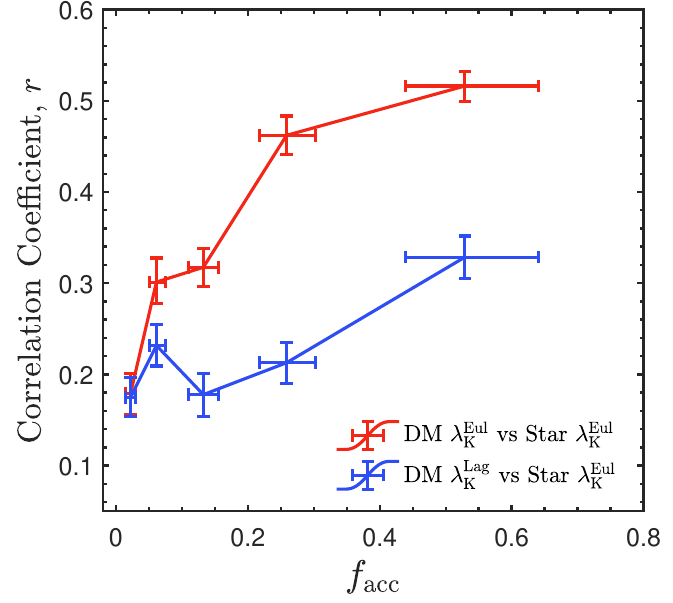}}
  \caption{Left: the number distributions of galaxy $ex$ $situ$ stellar mass fraction $f_{\rm acc}$ at 
  $z=0$. The black curve corresponds to the distribution of all galaxy samples. The disk-dominated, 
  intermediate-type, and spheroid-dominated galaxies are shown in blue, green, and red, respectively. 
  Middle: correlation between star-DM spin speed correlation and galaxy $ex$ 
  $situ$ stellar mass fraction $f_{\rm acc}$. Right: correlation between star-DM spin supportedness 
  correlation and galaxy $ex$ $situ$ stellar mass fraction $f_{\rm acc}$. Galaxies are binned every 800 
  samples by $f_{\rm acc}$ to calculate the Pearson correlation coefficients, $r$. The uncertainties 
  in the derived parameters are estimated by the bootstrap method, with the center, left/right 
  boundaries of the error bar representing the median, $25\%/75\%$ quartiles of the distributions. 
  The spin correlations between stars and Eulerian halos or Lagrangian protohalos are indicated by red and 
  blue curves, respectively.}
  \label{fig.6}
\end{figure*}

\begin{figure}[htbp]
  \centering    
  \subfigure{\includegraphics[width=0.9\columnwidth]{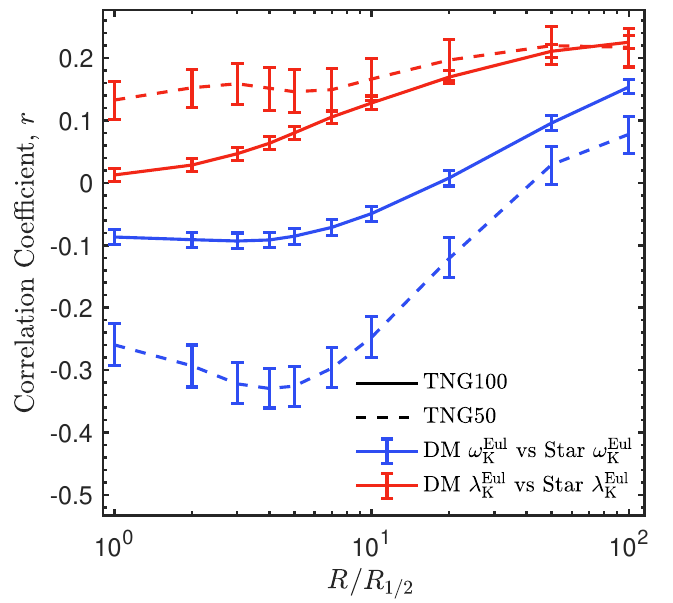}}
  \caption{Radial dependence of the star-DM spin magnitude correlation. The spin speed (blue line) and 
  supportedness (red line) of stars are measured within various radii $R$, normalized by the stellar half mass 
  radius $R_{1/2}$, to calculate the Pearson correlation coefficients with the total DM component. 
  The solid line represents the results in TNG100-1, while the dashed line shows the results for the galaxies 
  in the same mass range in TNG50-1. The error bars are similarly obtained in Fig. \ref{fig.6}.}
  \label{fig.7}
\end{figure}

In the left panel of Fig. \ref{fig.6}, we show the number distributions of galaxy $ex$ $situ$ stellar 
mass fractions, distinguishing between disc-dominated, intermediate-type and spheroid-dominated galaxies.
Most galaxies have much more $in$ $situ$ formed stars than $ex$ $situ$ accreted ones.
In addition, disc-dominated galaxies with $\kappa_{\rm rot}\geq0.7$ tend to have more $in$ $situ$ formed 
stars, while galaxies with a high $ex$ $situ$ stellar mass fraction are more likely to be spheroid-dominated 
with $\kappa_{\rm rot}<0.5$.

In the middle and right panels of Fig. \ref{fig.6}, we explore the relationship between the star-DM spin 
magnitude correlation and the $ex$ $situ$ stellar mass fraction. In Eulerian space, the spin speed and 
supportedness correlations between galaxy stellar component and their host DM halo increase with the $ex$ 
$situ$ stellar mass fraction, $f_{\rm acc}$. The Pearson correlation coefficient 
$r({\rm DM}\ \omega_{\rm K}^{\rm Eul}, {\rm Star}\ \omega_{\rm K}^{\rm Eul})$ of spin speed exhibits a 
rapid initial increase from $0.1$ to $0.5$ when $f_{\rm acc}<0.3$, followed by a slight further 
increment to its maximum value of $0.55$. The spin supportedness correlation shows a similar trend but is
slightly less sensitive to the increase in $ex$ $situ$ stellar mass fraction. 
In Fig. \ref{fig.7}, we show the radial dependence of the star-DM spin magnitude correlation. We 
measure the Pearson correlation coefficient $r$ between the spin magnitude of the total DM component and 
the spin magnitude of the stars enclosed within radius $R$. As shown by the solid lines in Fig. \ref{fig.7}, 
both the spin speed and supportedness correlations are positively correlated with the radius, indicating 
that the stars in the inner region of galaxies are poorly correlated with the DM component while the stars in 
the outer region are on the contrary.
These suggest that, in general, the spin magnitudes of $in$ $situ$ formed stars are primarily driven by the 
inner stellar disk, losing the correlation with their host DM halos. But the $ex$ $situ$ accreted stars are 
distributed in the outer region of galaxy-halo systems, which follow the DM component during merger events, 
resulting in a strong correlation. 

\begin{figure}[htbp]
  \centering    
  \subfigure{\includegraphics[width=0.49\columnwidth]{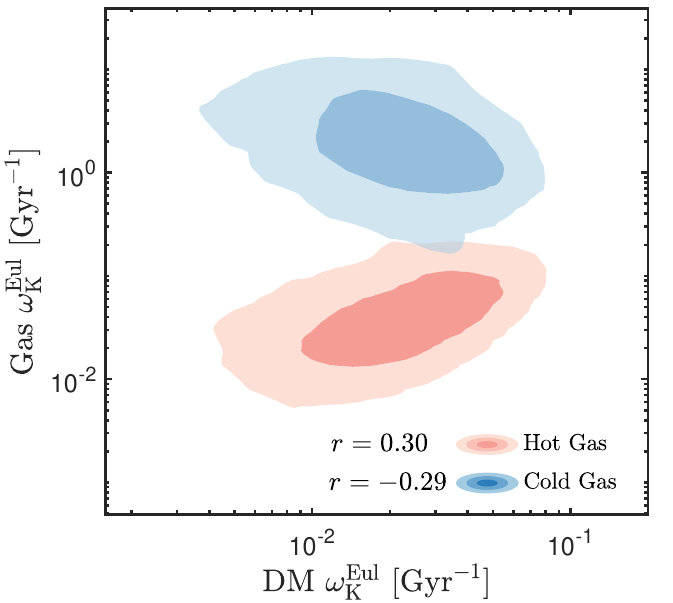}}
  \subfigure{\includegraphics[width=0.49\columnwidth]{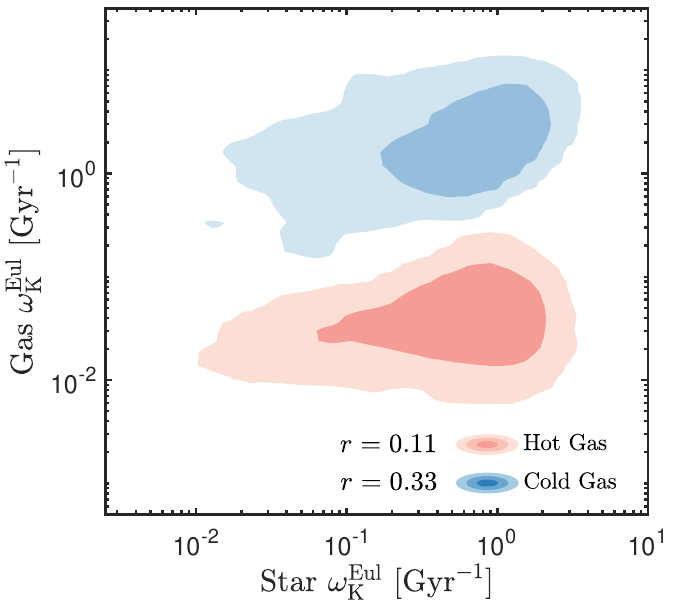}}
  \subfigure{\includegraphics[width=0.49\columnwidth]{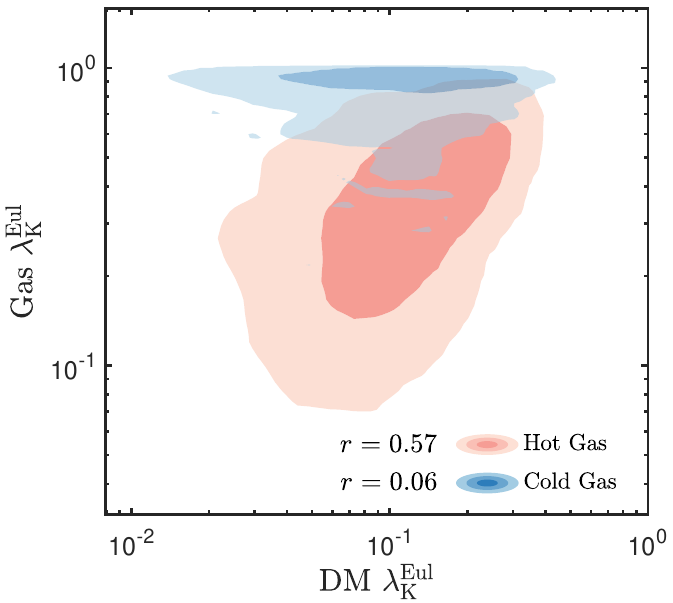}}
  \subfigure{\includegraphics[width=0.49\columnwidth]{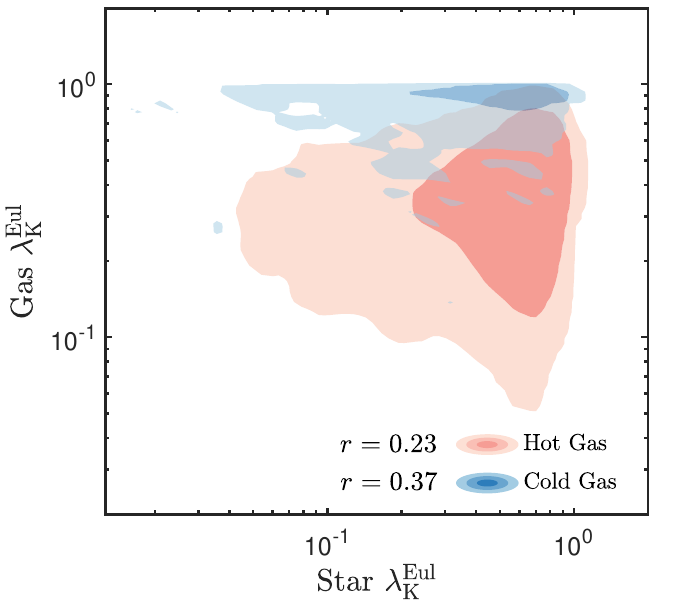}}
  \caption{Spin speed (upper row) and supportedness (lower row) correlations between hot or cold gas with DM 
  (left column) and stellar (right column) components. The red contour represents hot gas and blue represents 
  cold gas, with the inner and outer regions containing 68 and 95 percent of the galaxy population, respectively. 
  The Pearson correlation coefficients $r$ are indicated in each panel.}
  \label{fig.8}
\end{figure}

\begin{figure}[htbp]
  \centering    
  \subfigure{\includegraphics[width=0.90\columnwidth]{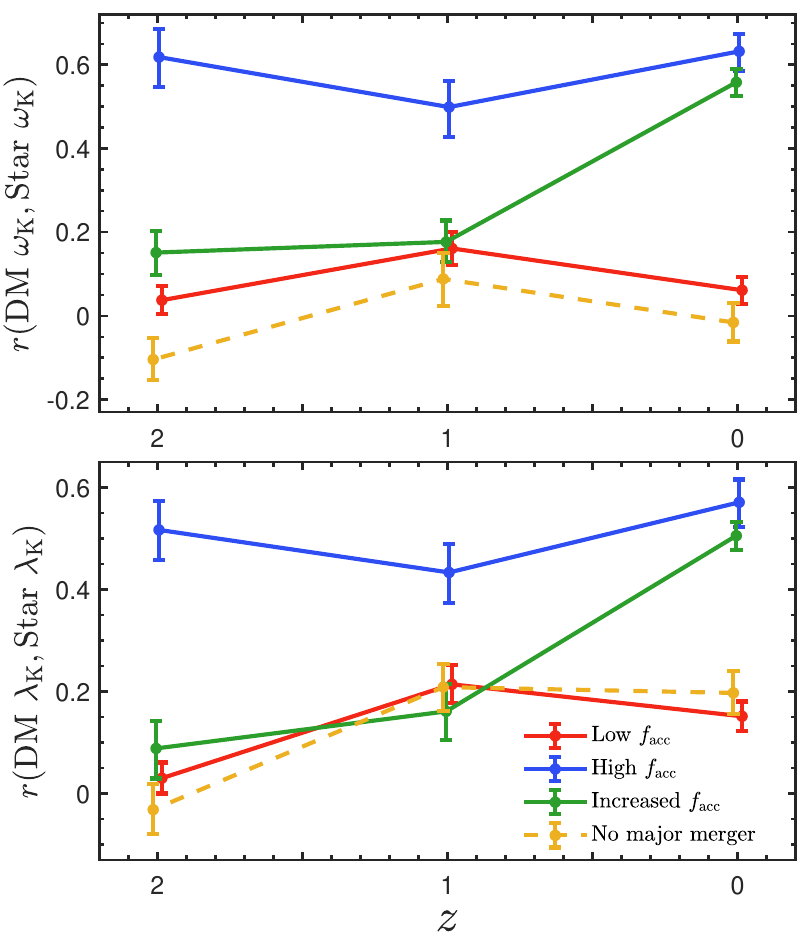}}
  \caption{The evolution of the star-DM spin speed (upper panel) and supportedness (lower panel) correlations from 
  redshift $z=2$ to $z=0$ for four typical types of galaxies. (i) Galaxies that always have low $ex$ $situ$ stellar mass 
  fractions $f_{\rm acc}$ (red solid line); (ii) galaxies that always have high $f_{\rm acc}$ (blue solid line); 
  (iii) galaxies with increased $f_{\rm acc}$ across time (green solid line); (iv) galaxies with no major merger event 
  yet (yellow dashed line). A detailed description of the classification criteria is provided in the main text. 
  The error bars are similarly obtained in Fig. \ref{fig.6}.}
  \label{fig.9}
\end{figure}

In Fig. \ref{fig.8}, we further show that both the spin speed and supportedness of stars are more correlated with 
those of cold gas, where we take the star-forming gas cells as cold and the rest as hot in TNG. The spin magnitude of dark 
matter indicates a strong correlation with hot gas, while there is a null or even negative correlation with cold gas. 
This consistently demonstrates that the spin magnitude of a subhalo/galaxy shows distinct distributions at different
radii, with the inner region reflecting the rotation of the gas disk and stellar disk and the outer region reflecting 
the assembly history. However, the spin direction correlation is less affected by the radii. Both the 
inner stellar disk and outer distributed stars have similar spin directions with their host halo, as shown in 
\citep{2023ApJ...943..128S}. In addition, \citep{2023ApJ...957...45X} suggested a similar dependence on $ex$ $situ$ 
stellar mass fraction $f_{\rm acc}$ for galaxy-halo alignment. 

Fig. \ref{fig.9} shows the evolution of the star-DM spin magnitude correlation from redshift $z=2$ to $z=0$. Here 
we trace back the progenitors of galaxies along their main progenitor branches in the merger trees. We select four typical 
types of galaxies from our samples: (i) galaxies that consistently have low $ex$ $situ$ stellar mass fractions $f_{\rm acc}$, 
that is, $f_{\rm acc}<0.05$, from redshift $z=2$ to $z=0$; (ii) galaxies that consistently have high $f_{\rm acc}$ 
($f_{\rm acc}>0.3$); (iii) galaxies with low $f_{\rm acc}$ at $z=2$, while increasing to high $f_{\rm acc}$ at $z=0$; (iv)
galaxies with no major merger event throughout the galaxy's history, provided by \citep{2017MNRAS.467.3083R,2023MNRAS.519.2199E} 
in the TNG simulation. Type (i) galaxies show a weak star-DM spin magnitude correlation across time, while type (ii) 
galaxies behave the opposite. Type (iii) galaxies accreted amounts of $ex$ $situ$ stars during their evolution history, 
resulting in an increased star-DM spin magnitude correlation. 
In particular, galaxies with no major merger event yet have a similar weak star-DM spin magnitude correlation with type (i) 
galaxies. Overall, the spin speed and supportedness correlations between the DM and stellar components at 
different redshifts show clear dependence on the $ex$ $situ$ stellar mass fractions. These results indicate that the 
connections between galaxies and their host halos are affected by their coevolutionary histories. 

In Lagrangian space, the spin magnitude correlation between the stellar component of galaxies
and their protohalos also shows a positive correlation with the $ex$ $situ$ stellar mass fraction. 
Compared with the correlation in Eulerian space, the Lagrangian correlation decreases but does not fully vanish, 
which indicates that the memory of the initial tidal fields of the stellar component is also not fully erased 
by the baryonic processes, and this memory is affected by the coevolutionary history of galaxy-halo systems.

\subsection{Connecting galaxy morphology to spin magnitude}
\begin{figure*}[htbp]
  \centering    
  \subfigure{\includegraphics[width=0.63\columnwidth]{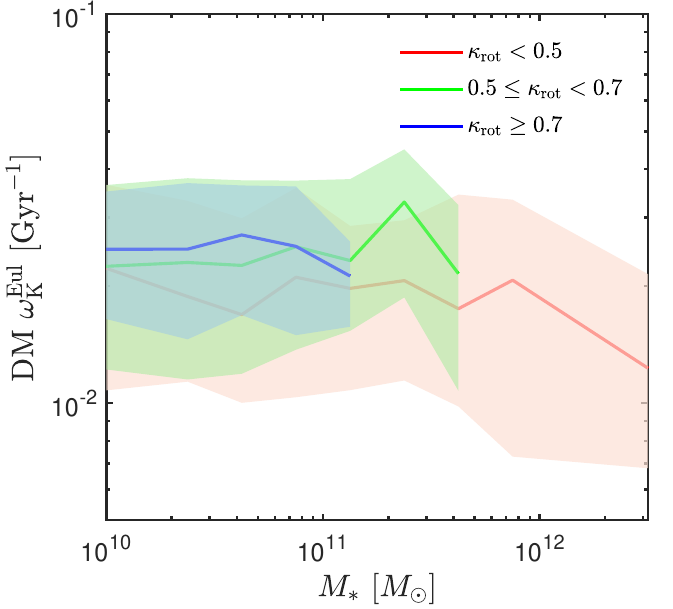}}
  \subfigure{\includegraphics[width=0.63\columnwidth]{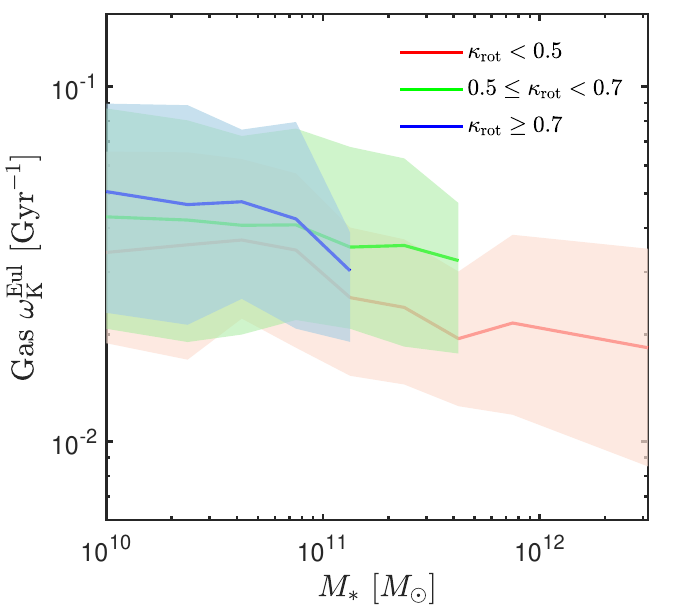}}
  \subfigure{\includegraphics[width=0.63\columnwidth]{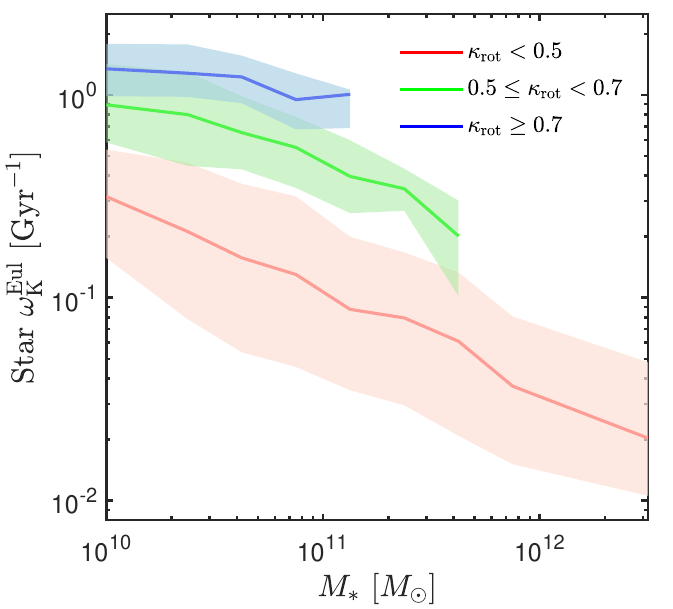}}
  \subfigure{\includegraphics[width=0.63\columnwidth]{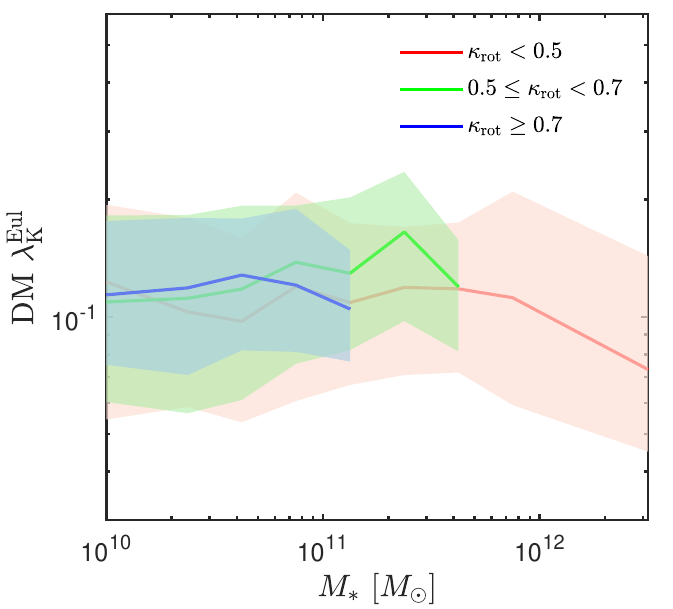}}
  \subfigure{\includegraphics[width=0.63\columnwidth]{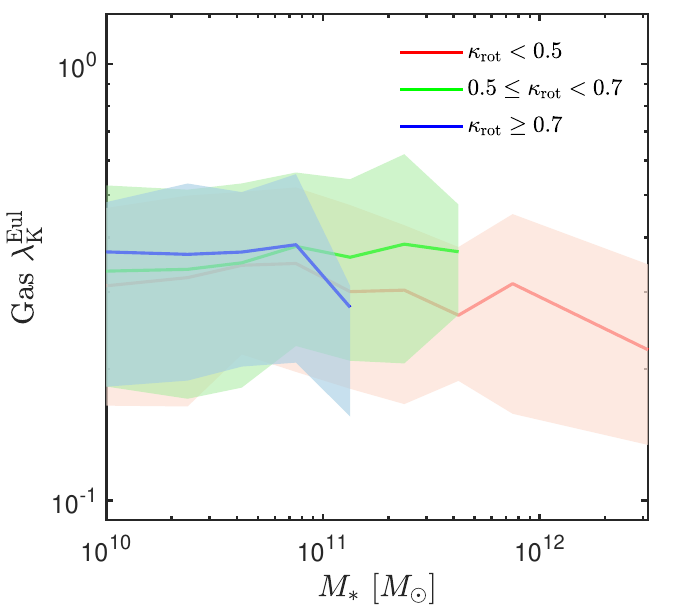}}
  \subfigure{\includegraphics[width=0.63\columnwidth]{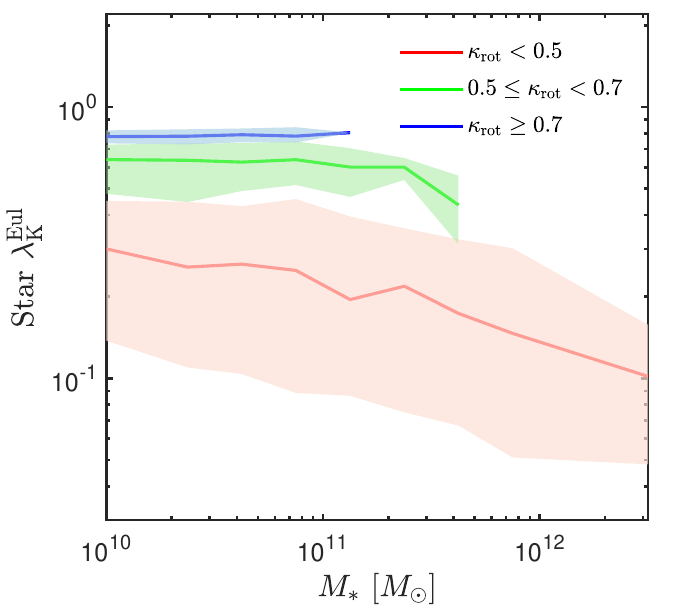}}
  \caption{The effect of DM (left column), gas (middle column), and stellar (right column) components' spin speed 
  (upper row) or supportedness (lower row) on galaxy morphology as a function of stellar mass.
  The blue, green, and red solid lines show the median trends for disc-dominated, intermediate-type, 
  and spheroid-dominated galaxies, respectively, while the shaded regions indicate the 16th to 84th 
  percentile ranges.}
  \label{fig.10}
\end{figure*}

Here, we directly study the connection between spin magnitude and galaxy morphology.
The left column of Fig. \ref{fig.10} suggests that disc-dominated galaxies with 
$\kappa_{\rm rot}\geq0.7$ have slightly higher spin speed $\omega_{\rm K}$ and supportedness $\lambda_{\rm K}$ 
for the DM component than spheroid-dominated galaxies, but the distinction is typically small. This indicates 
that galaxy morphology is not directly determined by the spin magnitude of its host halo, which is 
consistent with previous works \citep{2012MNRAS.423.1544S,2017MNRAS.467.3083R}. The results are 
similar for the gas components shown in the middle column. Naturally, the bottom-right panel shows that 
galaxy morphology is well distinguished by galaxy stellar spin supportedness as elaborated in Sec. 
\ref{sec.3.1}. Meanwhile, the top-right panel suggests that galaxy stellar spin speed is also strongly 
correlated with galaxy morphology. Disc-dominated galaxies have much higher stellar spin speeds and 
supportedness than spheroid-dominated ones.

\section{Conclusion and Discussion}\label{sec.conclu}
In this paper, by using the TNG100-1 simulation, we study the correlation of spin magnitudes 
between DM, gas, and stellar components of galaxy-halo systems, as well as their evolution throughout 
cosmic history. We conclude our new findings the following:
\begin{itemize}
  \item The DM, gas, and stellar components of galaxy-halo systems all have highly correlated spin speed 
  and supportedness in the Lagrangian and Eulerian spaces. These correlations are independent of galaxy 
  morphology, except for stars at low redshifts.
  \item Similar original mass distributions between DM and baryonic components lead to strong spin 
  speed and supportedness correlations between them in Lagrangian space. 
  These correlations are mostly conserved between gas and DM components at low redshifts. 
  Besides, the gas component still retains the memory of the original spin magnitude of its 
  host halo across the comic evolution, similar to that of the DM component but slightly weaker. 
  \item The $ex$ $situ$ stellar mass fraction $f_{\rm acc}$ is an important factor that could indicate
  the spin magnitude correlations between the galaxy stellar component and its host halo, as well as the 
  protohalo.
  The connection between galaxies and their host halos, as well as the memory of the initial 
  perturbations, are both correlated with their coevolutionary history.  
\end{itemize}

For a convergence check on resolution, we also test the result performed on a higher resolution simulation, 
TNG50-1. We select galaxies from the same mass range as our TNG100-1 galaxy samples. However, the number of galaxy 
samples in TNG50-1 is much smaller and contains fewer massive samples compared with TNG100-1 due to the smaller 
simulation box. In Fig. \ref{fig.7}, the dashed lines represent the result from TNG50-1, which shows a similar 
radial dependence of the star-DM spin magnitude correlation with TNG100-1. In addition, the rightmost data points 
indicate that the star-DM spin magnitude correlation is similar in TNG50-1 and 100-1, while the offsets in the 
inner region may arise from the particle resolution, galaxy sample size, and galaxy mass.

In our earlier paper \citep{2023ApJ...943..128S}, we found that the spin directions of DM halos and primordial spin 
modes can be well traced by baryonic matter (gas and stars) at low redshifts. Here we show that both the spin magnitude 
of DM halos and the initial spin magnitude can also be traced by the galaxy gas component, especially the hot gas 
component, and weakly by the stellar component.
The spin of the hot gas component can be observed via the kinetic 
Sunyaev-Zel'dovich (kSZ) effect \citep{1980MNRAS.190..413S}, which has been applied in many previous 
works \citep{2002ApJ...573...43C,2021MNRAS.504.4568M,2023MNRAS.519.1171Z,2023MNRAS.524.2262A}. 
Meanwhile, the initial galactic-halo spin can be predicted by tidal torque theory 
\citep{2002MNRAS.332..325P}.
This provides us with the possibility of using observable galaxy spin magnitude to constrain the 
cosmic initial conditions.

Nevertheless, the traceability of galaxy stellar component is largely affected by their assembly history. 
Galaxies with a high fraction of $in$ $situ$ formed stars are more likely to lose the spin magnitude 
correlation with their host halos as well as the initial perturbation. 
\citep{2021MNRAS.502.5480C,2022MNRAS.517.3459C} also found that the stellar angular momenta 
of galaxies tend to retain memory of the initial conditions just after mergers.
Meanwhile, \citep{2023ApJ...957...45X} reported that $f_{\rm acc}$ appears to be a fundamental parameter that 
determines the galaxy-halo alignment. These results indicate that the galaxy-halo correlations, both in 
shape and spin, are affected by the similarity of their evolution histories \citep{2022ApJ...936..119L}.
With the spin direction and magnitude of some specific galaxy-halo systems, e.g., the members of the local group, 
we can constrain their evolution histories. 

There are also some interesting questions that are worth further studying. First, whether the spin 
direction correlation between stellar and DM components is also affected by the $ex$ $situ$ stellar 
mass fraction $f_{\rm acc}$. 
Second, it would be interesting to compare the galaxy-halo correlation before and after an individual 
merger event and investigate how the galaxy loses the connection (shape and spin) to their host halo in 
and after star forming processes. Lastly, it is essential to quantitatively compare these results based on 
IllustrisTNG and other hydrodynamical simulations, as well as more convergence checks on resolution, 
such as the origin of the offsets of the correlations in the galaxy inner region. 
We leave these to future work.

\section*{Acknowledgements}
We thank Fangzhou Jiang, Yi Zheng, Min Du and Houjun Mo for valuable discussions and comments.
We also thank the anonymous referee for valuable suggestions.
This work is supported by National Science Foundation of China grant No. 12173030.
P.W. is sponsored by Shanghai Pujiang Program (No. 22PJ1415100).
The authors acknowledge the support by the China Manned Space Program through its Space 
Application System.
The IllustrisTNG simulations were run on the HazelHen Cray
XC40 supercomputer at the High Performance Computing Center Stuttgart 
(HLRS) as part of project GCS-ILLU of the Gauss
Centre for Supercomputing (GCS).

\bibliographystyle{h-physrev3}
\bibliography{mingjie_ref}

\end{document}